\documentclass{svproc}
\usepackage{cite}
\usepackage{amsmath,amssymb,amsfonts}
\usepackage{algorithmic}
\usepackage{graphicx}
\usepackage{textcomp}
\usepackage{hyperref}
\usepackage{xcolor}
\usepackage{amsmath}
\usepackage{graphicx}
\usepackage{array}
\usepackage{listings}
\usepackage{multirow}
\usepackage{xcolor}
\usepackage{listingsutf8}
\usepackage{soul}
\newcolumntype{P}[1]{>{\centering\arraybackslash}p{#1}}

\usepackage{listings}
\usepackage{multirow}
\usepackage{url}

\begin{document}
\mainmatter              

\title{Improving Semantic Similarity Measure within a Recommender System based-on RDF Graphs}
\titlerunning{Improving Similarity Measure within a RS based-on RDF Graphs}  
%
\author{LE Ngoc Luyen\inst{1,2}, Marie-Hélène ABEL\inst{1}, Philippe GOUSPILLOU\inst{2}
}
\institute{Université de technologie de Compiègne, CNRS, Heudiasyc (Heuristics and Diagnosis of Complex Systems), CS 60319 - 60203 Compiègne Cedex, France \and Vivocaz, 8 B Rue de la Gare, 02200, Mercin-et-Vaux, France}

%
%
%
\maketitle              

\begin{abstract}
In today's era of information explosion, more users are becoming more reliant upon recommender systems to have better advice, suggestions, or inspire them. The measure of the semantic relatedness or likeness between terms, words, or text data plays an important role in different applications dealing with textual data, as in a recommender system. Over the past few years, many ontologies have been developed and used as a form of structured representation of knowledge bases for information systems. The measure of semantic similarity from ontology has developed by several methods.  In this paper, we propose and carry on an approach for the improvement of semantic similarity calculations within a recommender system based-on RDF graphs.
\end{abstract}
\vspace{-1cm}
\section{Introduction}
\vspace{-0.25cm}
 With the development of the Internet, users face a large amount of information on e-commerce websites or mobile applications. On the other side, searching and matching algorithms have to deal with different structured, semi-structured, and unstructured textual data. The measure of semantic similarity allows comparing how close two terms or two entities are. Therefore, the exploitation of methods for semantic similarity measure becomes one of the means for improving these search and matching algorithms. In the context of a recommender system, semantic similarity measure can be applied in order to improve certain tasks such as searching, matching and ranking data. 

 A recommender system provides suggestions for items that are most likely of interest to the particular user. These suggestions can support users in various decision-making processes, for instance, which films to watch or which products to purchase. In order to give suggestions for items, the recommender systems attempt to collect characteristics about users and items by regarding user's preferences, items description, and behaviors. Thus, the measures of semantic similarity allows finding the most relevant items to the user.

By structuring and organizing a set of terms or concepts within a domain in a hierarchical way and by modeling the relationships between these sets of terms or concepts using a relation descriptor, an ontology allows to specify a standard conceptual vocabulary for representing entities in a particular domain \cite{rodriguez2003determining, 10.1145/2912845.2912869}. 
 In recent years, the use of ontologies has become more popular in recommender systems as well as in decision support systems for different tasks \cite{ibrahim2018ontology, obeid2018ontology, le2023constraint}. As a part of out work,  we are especially interested in calculating semantic similarity to improve the precision within a recommender system based-on RDF graphs. 


The rest of this article is organized as follows. First, we precise the principal problem that the paper focuses on and resolves. Then section 3 introduces works from the literature on which our approach is based.  Section 4 presents our main contributions on the construction of the recommender system exploiting the measure of similarity between sets of triplets. Before concluding, we test our work in section 5 from an experimental case dealing with the purchase/sale of used vehicles. Finally, we conclude and present the perspectives.

\vspace{-0.25cm}
\section{Problem statement}
\vspace{-0.25cm}
We are looking to improve the semantic similarity measure in order to obtain more accuracy on the items' recommendation lists to the user within the recommender system. In general, an ontology can be represented as a set of RDF triplets. An RDF triplet includes three components: subject, predicate and object. Common semantic similarity measure approaches based on an ontology have two weak points. The first weak point concerns the measure of similarity which is calculated either between objects, or between objects and predicates \cite{li2021ontology}. The object-based measure does not use the subject information, although it can contain contextual information from the triplet which is interesting for the comparison. The second weak point concerns the distinction of the type of objects: textual or numerical \cite{meym2016semantic}. The measure of similarity between numerical objects consists of a simple arithmetic calculation. The measure of similarity between textual objects is based on the frequency of the words composing the textual objects to be compared. This measure does not take into account the semantic dependence between these words. The latter can be a richness for comparison. 
\vspace{-0.25cm}
\section{Related Works}
\vspace{-0.15cm}
\subsection{Recommender Systems}
\vspace{-0.15cm}
The Recommender System (RS) is conventionally defined as an application that attempts to recommend the most relevant items to users by reasoning or predicting the user's preferences in an item based on related information about users, items, and the interactions between items and users \cite{LU201512, le2021towards}. In general, recommendation techniques can be classified into six main approaches: Demographic-based RSs, Content-based RSs, Collaboration Filtering-based RSs, Knowledge-based RSs, Context-aware RSs, and Hybrid RSs.

In several areas such as financial services, expensive luxury goods, real estate or automobiles, items are rarely purchased, and user reviews are often not available. In addition, item descriptions can be complex, and it is difficult to get a reasonable set of ratings that reflect users' history on a similar item. Therefore, demographic-based, content-based, collaborative filtering-based RSs are generally not well suited to domains where items possess the mentioned characteristics.

Recommender systems based on knowledge and contextual information represented by means of ontologies are proposed to address these challenges by explicitly soliciting user needs for these items and in-depth knowledge of the underlying domain for similarity measures and calculations of predictions \cite{jannach2010}.

To improve the quality of the recommendation, the similarity measures between items or user profiles in a recommender system plays a very important role. They make it possible to establish a list of recommendations taking into account the preferences of the users obtained following the declarations of the users or their interactions. We detail in the next section the measures of semantic similarity between items within a recommender system.
\vspace{-0.25cm}
\subsection{Semantic Similarity Measure}
\vspace{-0.3cm}
The advantages of using ontologies consist of the reuse of the knowledge base in various fields, traceability and the ability to use computation and application at a complex and large scale \cite{nguyen2011ontologies}.
Depending on the structure of the application context and its knowledge representation model, different similarity measures have been proposed. In general, these approaches can be categorized into four main strategies \cite{sanchez2012ontology, meym2016semantic}: (1) path-based, (2) feature-based, (3) information content-based, and (4) hybrid strategy.

With path-based semantic similarity measures, ontologies can be considered as a directed graph with nodes and links, in which classes or instances are interconnected mainly by means of hypernym and homonym relationships where the information is structured hierarchically using the `is-a' relationship \cite{meym2016semantic}. Thus, semantic similarities are calculated based on the distance between two classes or instances. 
The main advantage of this strategy is simplicity because it requires a low computational cost 
and does not require the detailed information of each class and instance \cite{li2021ontology}. The main drawback of this strategy concerns the degree of completeness, homogeneity, coverage and granularity of the relationships defined in the ontology \cite{sanchez2012ontology}.

When feature-based semantic similarity measures, classes and instances in ontologies are represented as a set of ontological features
\cite{sanchez2012ontology, meym2016semantic}. Commonalities between classes or instances are calculated based on their ontological feature set. 
The similarity evaluation can be performed using multiple coefficients on property sets such as the Jaccard index \cite{jaccard1901etude}, the Dice's coefficient \cite{dice1945}. The advantage of this strategy is that it evaluates both the commonalities and the differences of sets of compared properties which allow to exploit more semantic knowledge than the path-based approach. However, the limitation is that it is necessary to balance the contribution of each property by deciding the standardization and the weighting of the parameters on each property.

With semantic similarity measures based on information content, information content is used as a measure of information by associating probabilities of occurrence with each class or instance in the ontology and calculating the number of occurrences of these classes or instances \cite{sanchez2012ontology}. 
In this way, infrequent classes or instances become more informative than frequent classes or instances. A disadvantage of this strategy is that it requires large ontologies with a detailed taxonomic structure in order to properly differentiate between classes.

Beyond the measure of semantic similarities mentioned above, there are several approaches based on combinations of the three main strategies. Many works have combined the feature-based and path-based strategy \cite{hu2006semantic}.

In our work, we have chosen to work on the representation of an ontology by means of triplets. An RDF triplet has three components: subject, predicate and object. In particular, the subject can be the name of a class, or an instance. The predicate is the name of a property of a class or an instance. Object is a value of a property of the class or instance that can be separated into a literal or a name of another class or instance. 
The name of a class, an instance, or literals in triplets are expressed via a text that can include several words. In order to prepare their treatment, these textual contents are vectorized. We detail in the next section the methods we have studied for this purpose.
\vspace{-0.25cm}
\subsection{Vector Representation of Words}\label{vecteurdemot}
\vspace{-0.15cm}

Word vectorization allows to represent a word by a numeric feature vector and this vector describes the meaning of this word in its context. In general, several techniques are proposed to vectorize a word such as  Term Frequency-Inverse Document Frequency (TF-IDF) \cite{salton1986introduction}, or  Continuous Bag of Word (CBOW) and Skip Gram Model (Skip-Gram).

The TF-IDF is a statistical measure based on a corpus of documents\footnote{In the context of an ontology, a set of triples is equivalent to a document}. This technique assesses the relevance of a word to a document in a corpus of documents. 
The CBOW model constructs the vector representation of a word by predicting its occurrence and knowing the neighboring words. In another side, the Skip-Gram model constructs the vector representation of a word by predicting its context of occurrence. 

Word2vec is one of the most popular techniques for creating word embedding using a neural network architecture. It predicts words based on their context by combining the two models CBOW and Skip-gram \cite{mikolov2013efficient}. 
Several word embeddings are created using this model for different languages \cite{mikolov2013efficient}. Fauconnier \cite{fauconnier_2015}, and Hadi and his colleagues \cite{abdine2021evaluation} implement this model from French texts.

Word embedding trained with very large corpora allows to quickly obtain the vector representation of a word. In our work, we have chosen to calculate the measure of similarity between two textual terms taking into account the combination of CBOW and Skip-gram models. The similarity between two textual terms that consist of different words can take advantage of this form of representation in order to calculate the distance between them. In the next section, we detail our proposed approach to measure similarity within a RS.
\vspace{-0.25cm}
\section{Measure of Similarity within a Recommender System}
\vspace{-0.25cm}
\subsection{Recommender System for the Purchase/Sale of Used Vehicles}
\vspace{-0.15cm}
As part of our work, we are interested in the illustration of the semantic similarity measure on the RS based on the knowledge represented by means of ontologies in an e-commerce application for the sale/purchase of used vehicles. 

Knowledge base using for our RS represented by means of ontologies focuses on three main types: user profiles, item descriptions or item attributes, and interactions between users and items. First of all, user profiles include the user's personal information, their usage context, and their preferences about vehicle items. They can be organized and rewritten as triplets formally defined as follows:
\begin{equation}
	G_U = \{ a_1^u, a_2^u, ...,  a_n^u,  \}
\end{equation} where $a_i^u$ denotes the triplet $a_i^u = \langle subject_i, predicate_i, object_i \rangle$. In other words, the triplet $a_i^u$ can also be expressed as $\langle resource_i, property_i, state_i \rangle$ . For example, ``\textit{Louis likes the Tesla Model S car}''. This natural language expression can be represented through two different triplets: $\langle Louis, likes, the\_Tesla\_Model\_S\_car \rangle$, $\langle  The\_Tesla\_Model\_S\_car, is\_manufactured\_by,  Tesla\_Mo\-tors \rangle$. Then, vehicle descriptions can also be represented as a knowledge graph. They can be defined using the same approach:
\begin{equation}
	G_V = \{ a_1^v, a_2^v, ...,  a_m^v,  \}
\end{equation}where $a_j^v$ denotes the triplet $a_j^v = \langle subject_j, predicate_j, ob\-ject_j \rangle$ or $a_j^v = \langle resour\-ce_j, property_j, state_j \rangle$. Finally, when a user performs an interaction on vehicle description items by giving a rating, a comment or adding to a list of favorites, we mark these interactions to have an analysis of the intention and the behavior of the user in order to propose relevant vehicle item recommendations. Therefore, interactions are defined as a function with several parameters:
\begin{equation}
	RS : G_U \times G_V \times G_{C_1} \times ... \times G_{C_k} \rightarrow Interaction
\end{equation} where $G_U$ corresponds to the user, $G_V$ corresponds to the vehicle description item, $G_{C_h}$s corresponds to contextual information, for example: objectives, locations, times, resources \cite{Adomavicius2011}. 
Ontologies are developed to profile users and model vehicle description items \cite{le2021towards}. Based on these ontologies, RDF data is collected and stored in a searchable triplestore using SPARQL queries.
Rules can be defined to infer or filter items using inference ontologies. In this case, knowledge-based RS has the following four main tasks:
\begin{itemize}
	\item Receive and analyze user requests from the user interface.
	\item Build and perform queries on the knowledge base.
	\item Calculate semantic similarities between the vehicle description items, the user profile.
	\item Classify the items corresponding to the needs of the user.
\end{itemize}

Similarity measures between items or user profiles is an important task to generate the most relevant list of recommendations. 
Comparisons between two RDF triplets are often limited to common or non-common objects. The subject and predicate information can however also provide important information about the object itself and its comparison with other triplets. In the following section, we exploit information of triplets and calculate the semantic similarities between them in a knowledge base.
\vspace{-0.25cm}
\subsection{Semantic Similarity Measure between Triplets}
\vspace{-0.25cm}
We have chosen to define a hybrid approach that takes into account the combination of feature-based and content-based approaches to calculating similarities. The subject, predicate and object in triplet contain important information. A set of triplet allows to aggregate information from single triples. Therefore, the measure of semantic similarity between the sets of triplets must take into account all the triplets/elements in each set.

The measure of semantic similarity focuses on comparing two sets of triplets from all their elements by separating them into quantitative and qualitative information. On the one hand, the object comparison is performed using the property-based semantic similarity strategy. On the other hand, the comparison of subjects and predicates is performed by the semantic similarity strategy based on the content of information.

\subsubsection{Measure of Qualitative Information}
Qualitative information refers to words, labels used to describe classes, relationships, and annotations. In a triplet, the subject and the predicate express qualitative information. Objects can contain qualitative or quantitative information. For example, we have the following three triplets:
\begin{center}
\begin{tabular}{l}
$\langle ford\_focus\_4\_2018, has\_transmission, mechanical \rangle$\\
$\langle ford\_focus\_4\_2020, has\_transmission, mechanical \rangle$\\
$\langle citroen\_c5\_aircross, has\_transmission, mechanical \rangle$	\\
\end{tabular}
\end{center}

All components of these three triplets are qualitative. The subject information of three triplets can be used to contribute to the measure of similarity between them. In this section, we focus on measuring semantic similarity for  Qualitative Subjects, Predicates and Objects (QSPO). We propose the same formula for all three components to calculate similarity.

Let $a_{s1}$ and $a_{s2}$ be two QSPOs whose word vectors are $M_1 = \{w_{11}, w_{12}, ..., w_{1k}\}$ and $ M_2 = \{w_{21}, w_{22}, ..., w_{2l}\}$, their semantic similarity is defined as follows:
\begin{equation}
	Sim_1(a_{s1},a_{s2}) = \frac{\sum_{i=1}^{k} \bar{S}(w_{1i}, a_{s2}) + \sum_{j=1}^{l} \bar{S}(w_{2j}, a_{s1})}{k + l}
\end{equation}
where $\bar{S}(w, a_{s})$ denotes the semantic similarity of a word $w$ and a QSPO. The function $\bar{S}(w, a_{s})$ is formally calculated as follows:
\begin{equation}
	\bar{S}(w, a_{s}) = \max\limits_{w_i \in M} \bar{S}(w, w_i)
\end{equation} where $w_i \in M=\{w_1, w_2, ..., w_k\}$ is the word vector of $a_s$. Each word $w_i$ is represented by a numerical vector. One can use the techniques introduced in section \ref{vecteurdemot}. The TF-IDF word frequency-based approach facilitates obtaining the probability of a word in a set of triplets. However, the main disadvantage of this approach is that it cannot capture the semantic information of the word with the other words or the word order of the elements in the set of triplets because it creates the vector based on the frequency of the word in the set of triplets and the collection of sets of triplets. Therefore, we propose the use of CBOW and Skip-gram models with the implementation of Word2vec \cite{mikolov2013efficient, abdine2021evaluation} in order to overcome this weakness. We finally calculate the similarity between two words $w_i$, $w_j$ by cosine similarity: $\bar{S}(w_i, w_j) = \frac{w_i . w_j}{\|w_i\|\|w_j\|}$.
\vspace{-0.25cm}
\subsubsection{Measure of Quantitative Information}
Quantitative information is numerical information that is used to express nominal, ordinal, interval, or ratio information. In a triplet, the object often uses this form of information to manifest property information for classes, concepts of the ontology. For example, we have the following triplets: 
\begin{center}
	\begin{tabular}{l}
$\langle ford\_focus\_4\_2018, has\_number\_of\_mileage, 107351\rangle$\\
$\langle ford\_focus\_4\_2020, has\_number\_of\_mileage, 25040 \rangle$\\
$\langle citroen\_c5\_aircross, has\_number\_of\_mileage, 48369 \rangle$ \\
\end{tabular}
\end{center}
The objects of these triplets are numeric values. The comparison between numbers is done simply by measuring the distance. In order to compare two different objects, we use the Euclidean distance between two objects. Thus, the smaller the difference between two objects, the higher similarity between them. Let $a_{o1}$ and $a_{o1}$ be two objects whose vectors are $a_{o1} = \{o_{11}, o_{12}, ..., o_{1k}\}$ and $a_{o2} =\{o_{21}, o_{22}, ..., o_{2k}\}$, their semantic similarity is defined as follows:
\begin{equation}
	Sim_2(a_{o1},a_{o2}) = \frac{1}{1 + \sqrt{\sum_{i=0}^{k}(o_{1i} - o_{2i})^2}}
\end{equation}
\vspace{-0.5cm}
\subsubsection{Measure of Triplets}
The comparison of two triplets $a_1=\langle a_{s1}, a_{p1}, a_{o1} \rangle$ and $a_2=\langle a_{s2}, a_{p2}, a_{o2} \rangle$ is performed according to the type of information of the objects in the triplets. If the object contains qualitative information, the semantic similarity between $a_1$ and $a_2$ is defined as follows:
\begin{equation}
	Sim_{I} (a_1,a_2) = \frac{1}{N}\sum_{i \in P, \omega \in Q} \omega \times Sim_1(a_{i1},a_{i2}) 
\end{equation} where $P = \{s, p, o\}$ corresponds to the information of $subject$, $predicate$, and $object$ as a vector of words. $Q = \{\alpha, \beta, \gamma\}$ is the respective weights for the triplet components. $N$ is the number of triplet components.

Moreover, if the object contains quantitative information, the semantic similarity measure of the triplets $a_1$ and $a_2$ is defined as follows:
\begin{equation}
	\begin{split}
		Sim_{II} (a_1,a_2) =  \frac{1}{N} (\sum_{i \in P, \omega \in Q} \omega \times Sim_1(a_{i1},a_{i2}) + \gamma \times Sim_2(a_{o1},a_{o2}))
	\end{split}
\end{equation}
where $P = \{s, p\}$ corresponds to the information of $subject$ and $predicate$ in the form of a vector of words. $Q = \{\alpha, \beta\}$ represents the respective weights of the subject and the predicate. And $\gamma$ is the weight for the object.

Therefore, the semantic similarity of two sets of triplets $G_1=\{a_1, a_2, ..., a_g\}$ and $G_2=\{a_1, a_2, ..., a_g\}$ is calculated on the basis for similarity comparison of each simple triplet as follows:
\begin{equation}
	\begin{split}
		Sim(G_1, G_2) = \frac{1}{L}(\sum_{i=0}^{L} Sim_{I}(a_{1i},a_{2i})) \;+  \frac{1}{H}(\sum_{j=0}^{H}Sim_{II}(a_{1j},a_{2j})) 
	\end{split}
\end{equation} where $L$ is the number of triplets that contains the qualitative objects. $H$ is the number of triplets that contains the quantitative objects.
\vspace{-0.25cm}
\section{Experiments}
\begin{figure}[h!]
	\vspace{-0.25cm}
	\begin{center}
		\includegraphics[width=0.90\textwidth]{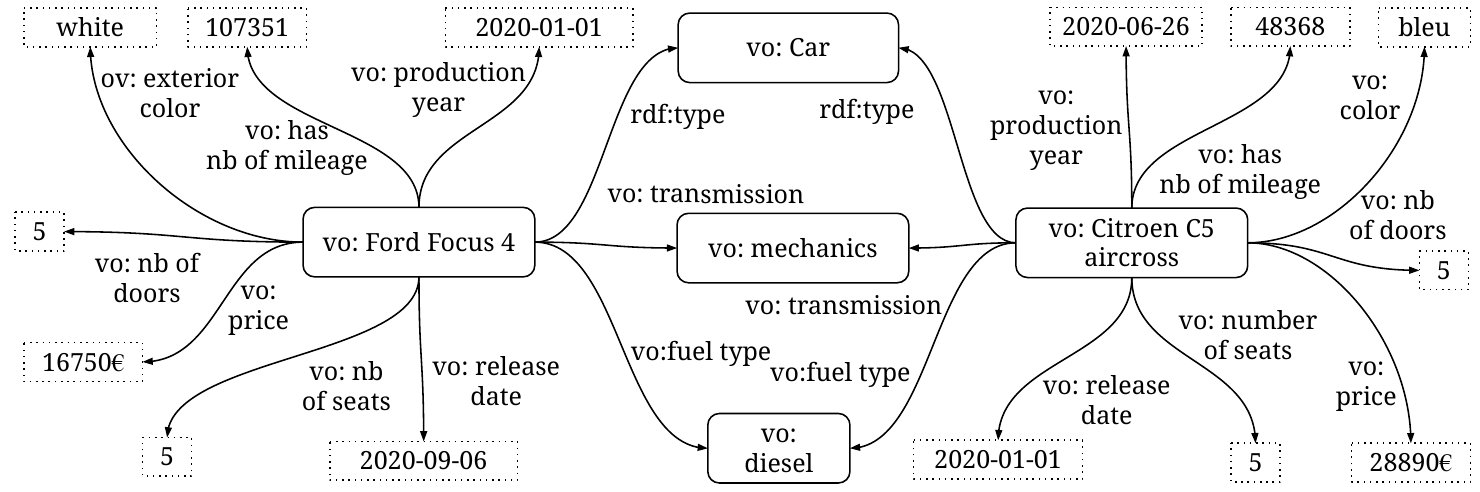}
	\end{center}
\vspace{-0.5cm}
	\caption{Triplet data visualized by a graph of two different vehicles (\textit{vo} notes for the Vehicle Ontology)}
	\label{fig02}
	\vspace{-0.2cm}
\end{figure}

\vspace{-0.25cm}
In this section we test our approach in the case of a vehicle purchase/sale application. We thus measure the semantic similarity between two sets of triplets each representing a vehicle. By using ontology, we can reconstruct the knowledge base of a domain in a form that is readable by machines as well as humans. From the vehicle ontologies developed in the work \cite{le2021towards, lengochal03675591}, we realize a collection of class instances and their relationships to create an RDF dataset. Figure \ref{fig02} illustrates  in a simple way two sets of triplets representing two vehicles. The dataset contains approximately 1000 used vehicles with its different features, characteristics.

The transformation of words into vector representation is achieved by using pretrained word embeddings for French developed by Hadi and his colleagues \cite{abdine2021evaluation}. We chose to employ the CBOW and Skip-gram models instead of TF-IDF model because the problem concerns capturing semantic information which is almost impossible on the TF-IDF model. 
\begin{figure}[h!]
	\begin{center}
		\includegraphics[width=0.90\textwidth]{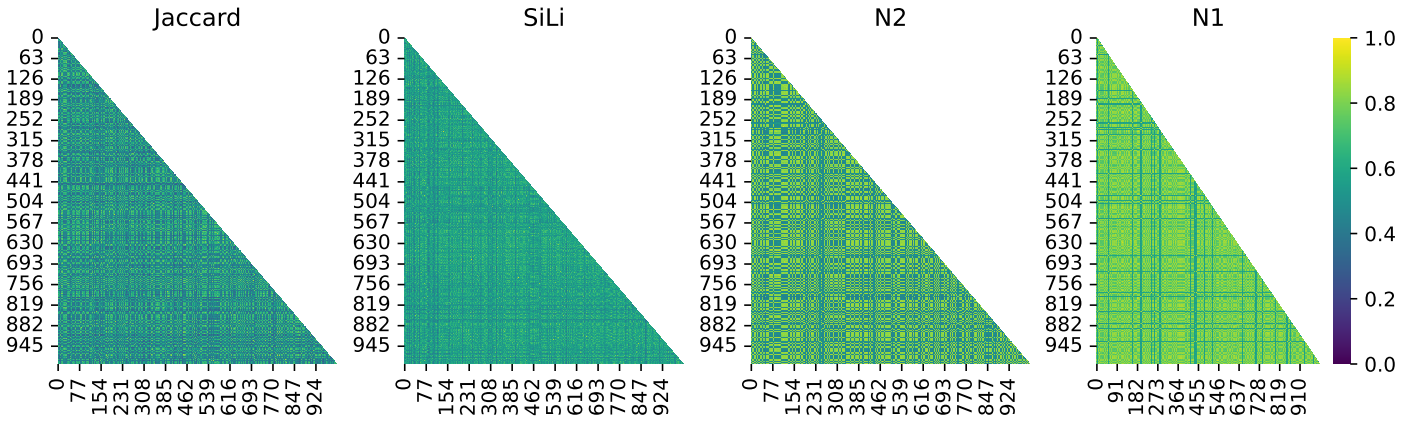}
	\end{center}
	\vspace{-0.5cm}
	\caption{Heat map correlation among similarity measures across 1000 used vehicles for four approaches}
	\label{fig03}
	\vspace{-0.5cm}
\end{figure}

Based on the instances collected, we carry out experiments and evaluations on the following four approaches:
\begin{enumerate}
	\item \textbf{Jaccard}: the approach based on Jaccard index that measures similarities between two triplet sets \cite{fletcher2018comparing}.
	\item \textbf{SiLi}: the approach proposed by Siying Li and her colleagues \cite{li2021ontology}, this hybrid approach combines the strategy based on the content of information and based on features but only considers the objects and the predicates of the triplets.
	\item \textbf{N2}: our approach with the use of the TF-IDF model to vectorize qualitative information
	\item \textbf{N1}: our main proposed approach with the use of the Word2vec model \cite{abdine2021evaluation} to vectorize qualitative information.	
\end{enumerate}


\begin{figure}[h!]
	\vspace{-1.0cm}
	\begin{center}
		\includegraphics[width=0.86\textwidth]{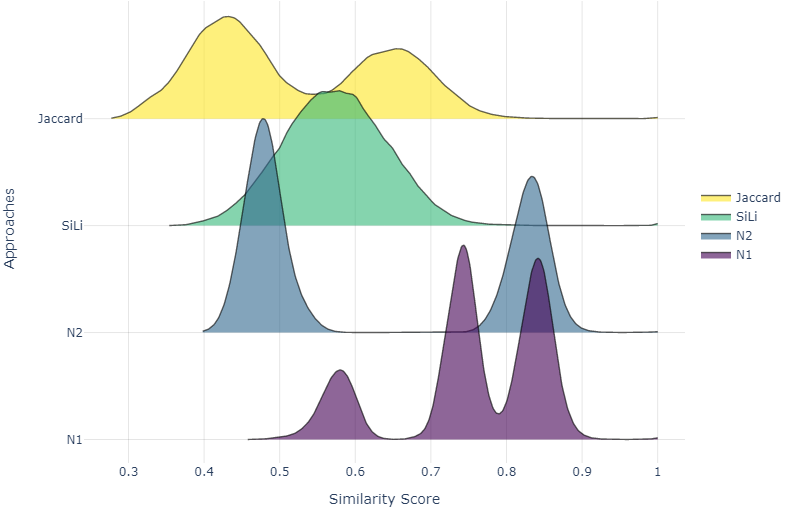}
	\end{center}
	\vspace{-0.5cm}
	\caption{Histogram representing the similarity score distribution of the four approaches}
	\label{fig04}
	\vspace{-0.5cm}
\end{figure}
We experiment with the four approaches by measuring the similarity scores for each RDF instance of used vehicles with all others. With RDF graphs of 1000 used vehicle instances, we have a total of 100000 similarity scores. In general, the approach \textbf{N1} yields better results when compared to the approaches \textbf{N2}, \textbf{Sili}, and \textbf{Jaccard} in terms of similarity score through some aspects of the results that are represented by heat map (figure \ref{fig03}) and histogram chart(figure \ref{fig04}). The yellow color distribution on the heat map illustrates the higher similarity score of the approach \textbf{N1}  than the other approaches and is evenly distributed throughout the map. Likewise, the histogram chart also shows that the distribution of similarity score of the approach \textbf{N1} is higher than the others. 

A deep dive into the results obtained, we arrive at several conclusions. First, our approach \textbf{N1} gives the result of the calculation of the similarity between the vehicles higher than the other approaches with 82.4 percent highest scores. Second, our approach using the TF-IDF \textbf{N2} technique for vector word representation obtained the lower results by comparing with the approach \textbf{N1}. This is explained by the ability to capture contextual and semantic information of the Word2vec approach which is better than that of the TF-IDF approach.

Experiments show that our approach \textbf{N1} obtained good results for similarity measures between the sets of triples. The use of the subject in the comparison allows to add information to the measure of similarity of a triplet. Also, the distinction between textual and numerical content allows to apply the appropriate formula according to the type of content. The sum of the two calculations represents the measured similarity. Taking into account this distinction, the contextual triplets and the measure from the textual contents enriched with the semantic dependencies between the words constituting the text, the similarity obtained is more precise than those encountered in the literature \cite{sanchez2012ontology, meym2016semantic, li2021ontology}.
\vspace{-0.25cm}
\section{Conclusion and Perspectives}
\vspace{-0.25cm}
Measuring semantic similarity based on ontology is an important task in proposing a list of relevant recommendations to a user. In this article, we propose a hybrid strategy that combines feature-based strategy and content-based information. The two weak points in the problem statement section are intervened by our semantic similarity measure approach: (1) the three components of a triplet are considered in the similarity measure in order not to lose information, and (2) The distinction of data type, textual or numeric, allows to carry out an adapted and more precise measure. We carried out an experiment of our approach and compared with three other similarity measures. The results obtained show its interest. Now we need to continue our work and carry out other tests on different corpora and different applications. We must concede that the words not considered in the trained corpus pose a problem. In perspective, the research for treating and cleaning a corpus of the vehicle domain as well as applying ontologies of the domain in order to improve the precision of the recommender systems could be promising works in the future.
\vspace{-0.25cm}
\section*{Acknowledgment}
\vspace{-0.25cm}
This work was funded by the French Research Agency (ANR) and by the company Vivocaz under the project France Relance - preservation of R\&D employment (ANR-21-PRRD-0072-01).
\vspace{-0.25cm}
%
%
\bibliographystyle{plain}
\bibliography{bibliotheque}

\begin{thebibliography}{10}

\bibitem{abdine2021evaluation}
Hadi Abdine, Christos Xypolopoulos, Moussa~Kamal Eddine, and Michalis
  Vazirgiannis.
\newblock Evaluation of word embeddings from large-scale french web content.
\newblock 2021.

\bibitem{Adomavicius2011}
Gediminas Adomavicius and Alexander Tuzhilin.
\newblock {\em Context-Aware Recommender Systems}, pages 217--253.
\newblock Springer US, Boston, MA, 2011.

\bibitem{dice1945}
Lee~R Dice.
\newblock Measures of the amount of ecologic association between species.
\newblock {\em Ecology}, 26(3):297--302, 1945.

\bibitem{fauconnier_2015}
Jean-Philippe Fauconnier.
\newblock French word embeddings, 2015.

\bibitem{fletcher2018comparing}
Sam Fletcher, Md~Zahidul Islam, et~al.
\newblock Comparing sets of patterns with the jaccard index.
\newblock {\em Australasian Journal of Information Systems}, 22, 2018.

\bibitem{hu2006semantic}
Bo~Hu, Yannis Kalfoglou, Harith Alani, David Dupplaw, Paul Lewis, and Nigel
  Shadbolt.
\newblock Semantic metrics.
\newblock In {\em International Conference on Knowledge Engineering and
  Knowledge Management}, pages 166--181. Springer, 2006.

\bibitem{ibrahim2018ontology}
Mohammed~E Ibrahim, Yanyan Yang, David~L Ndzi, Guangguang Yang, and Murtadha
  Al-Maliki.
\newblock Ontology-based personalized course recommendation framework.
\newblock {\em IEEE Access}, 7:5180--5199, 2018.

\bibitem{jaccard1901etude}
Paul Jaccard.
\newblock {\'E}tude comparative de la distribution florale dans une portion des
  alpes et des jura.
\newblock {\em Bull Soc Vaudoise Sci Nat}, 37:547--579, 1901.

\bibitem{jannach2010}
Dietmar Jannach, Markus Zanker, Alexander Felfernig, and Gerhard Friedrich.
\newblock {\em Knowledge-based recommendation}, page 81–123.
\newblock Cambridge University Press, 2010.

\bibitem{lengochal03675591}
Ngoc~Luyen Le, Marie-H{\'e}l{\`e}ne Abel, and Philippe Gouspillou.
\newblock {Apport des ontologies pour le calcul de la similarit{\'e}
  s{\'e}mantique au sein d'un syst{\`e}me de recommandation}.
\newblock In {\em {Ing{\'e}nierie des Connaissances (Ev{\`e}nement affili{\'e}
  {\`a} PFIA'22 Plate-Forme Intelligence Artificielle)}}, Saint-{\'E}tienne,
  France, June 2022.

\bibitem{le2021towards}
Ngoc~Luyen Le, Marie-H{\'e}l{\`e}ne Abel, and Philippe Gouspillou.
\newblock Towards an ontology-based recommender system for the vehicle sales
  area.
\newblock In {\em Progresses in Artificial Intelligence {\&} Robotics:
  Algorithms {\&} Applications}, pages 126--136, Cham, 2022. Springer
  International Publishing.

\bibitem{10.1145/2912845.2912869}
Ngoc~Luyen LE, Anne Tireau, Aravind Venkatesan, Pascal Neveu, and Pierre
  Larmande.
\newblock Development of a knowledge system for big data: Case study to plant
  phenotyping data.
\newblock In {\em Proceedings of the 6th International Conference on Web
  Intelligence, Mining and Semantics}, WIMS '16, 2016.

\bibitem{le2023constraint}
Ngoc~Luyen L{\^e}, Jinfeng Zhong, Elsa Negre, and Marie-H{\'e}l{\`e}ne Abel.
\newblock Constraint-based recommender system for crisis management
  simulations.
\newblock 2023.

\bibitem{li2021ontology}
Siying Li, Marie-H{\'e}l{\`e}ne Abel, and Elsa Negre.
\newblock Ontology-based semantic similarity in generating context-aware
  collaborator recommendations.
\newblock In {\em 2021 IEEE 24th International Conference on Computer Supported
  Cooperative Work in Design (CSCWD)}, pages 751--756. IEEE, 2021.

\bibitem{LU201512}
Jie Lu, Dianshuang Wu, Mingsong Mao, Wei Wang, and Guangquan Zhang.
\newblock Recommender system application developments: A survey.
\newblock {\em Decision Support Systems}, 74:12--32, 2015.

\bibitem{meym2016semantic}
Rouzbeh Meymandpour and Joseph~G Davis.
\newblock A semantic similarity measure for linked data: An information
  content-based approach.
\newblock {\em Knowledge-Based Systems}, 109:276--293, 2016.

\bibitem{mikolov2013efficient}
Tomas Mikolov, Kai Chen, Greg Corrado, and Jeffrey Dean.
\newblock Efficient estimation of word representations in vector space.
\newblock {\em arXiv preprint arXiv:1301.3781}, 2013.

\bibitem{nguyen2011ontologies}
Van Nguyen.
\newblock Ontologies and information systems: a literature survey.
\newblock 2011.

\bibitem{obeid2018ontology}
Charbel Obeid, Inaya Lahoud, Hicham El~Khoury, and Pierre-Antoine Champin.
\newblock Ontology-based recommender system in higher education.
\newblock In {\em Companion Proceedings of the The Web Conference 2018}, pages
  1031--1034, 2018.

\bibitem{rodriguez2003determining}
M~Andrea Rodriguez and Max~J. Egenhofer.
\newblock Determining semantic similarity among entity classes from different
  ontologies.
\newblock {\em IEEE transactions on knowledge and data engineering},
  15(2):442--456, 2003.

\bibitem{salton1986introduction}
Gerard Salton and J~McGill.
\newblock Introduction to modern information retrieval.
\newblock 1986.

\bibitem{sanchez2012ontology}
David S{\'a}nchez, Montserrat Batet, David Isern, and Aida Valls.
\newblock Ontology-based semantic similarity: A new feature-based approach.
\newblock {\em Expert systems with applications}, 39(9):7718--7728, 2012.

\end{thebibliography}
\end{document}